\documentclass[%
 reprint,
superscriptaddress,
 amsmath,amssymb,
 aps,
]{revtex4-2}

\usepackage{graphicx}
\usepackage{dcolumn}
\usepackage{bm}
\usepackage{hyperref}
\usepackage{ulem}

\usepackage{color}

\begin{document}


\title{Single crystal growth of iridates without platinum impurities}

\author{Jimin Kim}
\email{jiminkim05@ibs.re.kr}
\affiliation{Center for Artificial Low Dimensional Electronic Systems, Institute for Basic Science (IBS), 77 Cheongam-Ro, Pohang 37673, South Korea}
\author{Hoon Kim}
\affiliation{Center for Artificial Low Dimensional Electronic Systems, Institute for Basic Science (IBS), 77 Cheongam-Ro, Pohang 37673, South Korea}
\affiliation{Department of Physics, Pohang University of Science and Technology, Pohang 37673, South Korea}
\author{Hyun-Woo J. Kim}
\affiliation{Center for Artificial Low Dimensional Electronic Systems, Institute for Basic Science (IBS), 77 Cheongam-Ro, Pohang 37673, South Korea}
\affiliation{Department of Physics, Pohang University of Science and Technology, Pohang 37673, South Korea}
\author{Sunwook Park}
\affiliation{Center for Artificial Low Dimensional Electronic Systems, Institute for Basic Science (IBS), 77 Cheongam-Ro, Pohang 37673, South Korea}
\affiliation{Department of Physics, Pohang University of Science and Technology, Pohang 37673, South Korea}
\author{Jin-Kwang Kim}
\affiliation{Center for Artificial Low Dimensional Electronic Systems, Institute for Basic Science (IBS), 77 Cheongam-Ro, Pohang 37673, South Korea}
\affiliation{Department of Physics, Pohang University of Science and Technology, Pohang 37673, South Korea}
\author{Junyoung Kwon}
\affiliation{Center for Artificial Low Dimensional Electronic Systems, Institute for Basic Science (IBS), 77 Cheongam-Ro, Pohang 37673, South Korea}
\affiliation{Department of Physics, Pohang University of Science and Technology, Pohang 37673, South Korea}
\author{Jungho Kim}
\affiliation{Advanced Photon Source, Argonne National Laboratory, Argonne, Illinois 60439, USA}
\author{Hyeong Woo Seo}
\affiliation{Center for Artificial Low Dimensional Electronic Systems, Institute for Basic Science (IBS), 77 Cheongam-Ro, Pohang 37673, South Korea}
\affiliation{Department of Physics, Pohang University of Science and Technology, Pohang 37673, South Korea}
\author{Jun Sung Kim}
\affiliation{Center for Artificial Low Dimensional Electronic Systems, Institute for Basic Science (IBS), 77 Cheongam-Ro, Pohang 37673, South Korea}
\affiliation{Department of Physics, Pohang University of Science and Technology, Pohang 37673, South Korea}
\author{B. J. Kim}
\email{bjkim6@postech.ac.kr}
\affiliation{Center for Artificial Low Dimensional Electronic Systems, Institute for Basic Science (IBS), 77 Cheongam-Ro, Pohang 37673, South Korea}
\affiliation{Department of Physics, Pohang University of Science and Technology, Pohang 37673, South Korea}


\date{\today}

\begin{abstract}
Iridates have attracted much interest in the last decade for their novel magnetism emerging in the limit of strong spin-orbit coupling and possible unconventional superconductivity. A standard for growing iridate single crystals has been the flux method using platinum crucibles. Here, we show that this widely used method compromises the sample quality by inclusion of platinum impurities. We find that Sr$_2$IrO$_4$ single crystals grown in iridium crucibles show remarkable differences from those grown in platinum crucibles in their sample characterizations using Raman spectroscopy, resistivity, magnetization, optical third harmonic generation, resonant X-ray diffraction, and resonant inelastic X-ray scattering measurements. In particular, we show that  several peaks of sizable intensities disappear in the Raman spectra of samples free of platinum impurities, and a significantly larger activation energy is extracted from the resistivity data compared to previously reported values. Furthermore, we find no evidence of the previously reported glide symmetry breaking structural distortions and confirm the $I4_1/acd$ space group of the lattice symmetry. Although the platinum impurities are not apparent in the magnetic properties and thus went unnoticed in the stoichiometric insulating phase for a long time, their effects can be much more detrimental to transport properties in chemically doped compounds. Therefore, our result suggests using growth methods that avoid platinum impurities for an investigation of intrinsic physical properties of iridates, and possible superconducting phases.

\end{abstract}


\maketitle


 
Over the last decades, iridates have been a central playground for the study of Mott physics in the limit of strong spin-orbit coupling (SOC), owing to their novel electronic and magnetic properties arising from spin-orbit entangled $J_{\textrm{eff}}$=1/2 pseudospins\cite{Kim08, Kim09, Jack09, Bale10, Witc13}. In particular, Sr$_2$IrO$_4$, a single-layer member of the Ruddlesden-Popper (RP) series iridates\cite{Craw94,Cao98}, has attracted wide interest as a prototypical spin-orbit Mott insulator, in which the pseudospins, despite the strong SOC, realize a Heisenberg antiferromagnet on a quasi-two-dimensional square lattice, in striking similarity with parent insulators of cuprate high temperature superconductors\cite{Wang11,Kim12,Bert19}. Indeed, theoretical studies using Monte Carlo\cite{Wata13}, functional renormalization group\cite{Yang14}, and dynamical mean field theory simulations\cite{Meng14} have indicated possible unconventional superconductivity upon carrier doping, while experimental studies using surface $in\:situ$ doping combined with angle-resolved photoemission spectroscopy (ARPES)\cite{Kim14,Kim16} and scanning tunneling spectroscopy (STS)\cite{Yan15} have found tantalizing evidence of unconventional superconductivity: the Fermi surface breaks up into disconnected arcs with pseudogap opening, which at low temperatures evolve to a gap structure consistent with $d$-wave superconductivity.  

However, superconductivity in the square-lattice iridates remains elusive to date. A prerequisite for the search for superconductivity is to grow bulk-doped samples with reasonably high quality, which at high enough doping concentration should exhibit a metallic phase with a well-defined Fermi surface. Although 
ARPES studies on La-doped Sr$_2$IrO$_4$ have partially reproduced the results from the $in\:situ$ doping study\cite{DeLa15}, a coherent quasiparticle has, to the best of our knowledge, never been observed in chemically doped samples\cite{Cao16}. 
Even for the stoichiometric parent insulator, STM topograph images taken on the surface of single\cite{Sun21} and bilayer\cite{Zhao21} iridates show a large density of unidentified impurities. Furthermore, glide-symmetry-breaking structural distortions found from neutron diffraction\cite{Dhit13, Ye13, Ye15} and optical third harmonic generation (THG) measurements\cite{Torc15} are reported to be sample dependent\cite{Seyl20}.  
These results raise quality issues in currently prevailing samples and call for a critical reexamination of the growth method\cite{Sung16, Mann20} to obtain high-quality single crystals and reveal the intrinsic physical properties of iridates.


\begin{figure*}[t]
\includegraphics{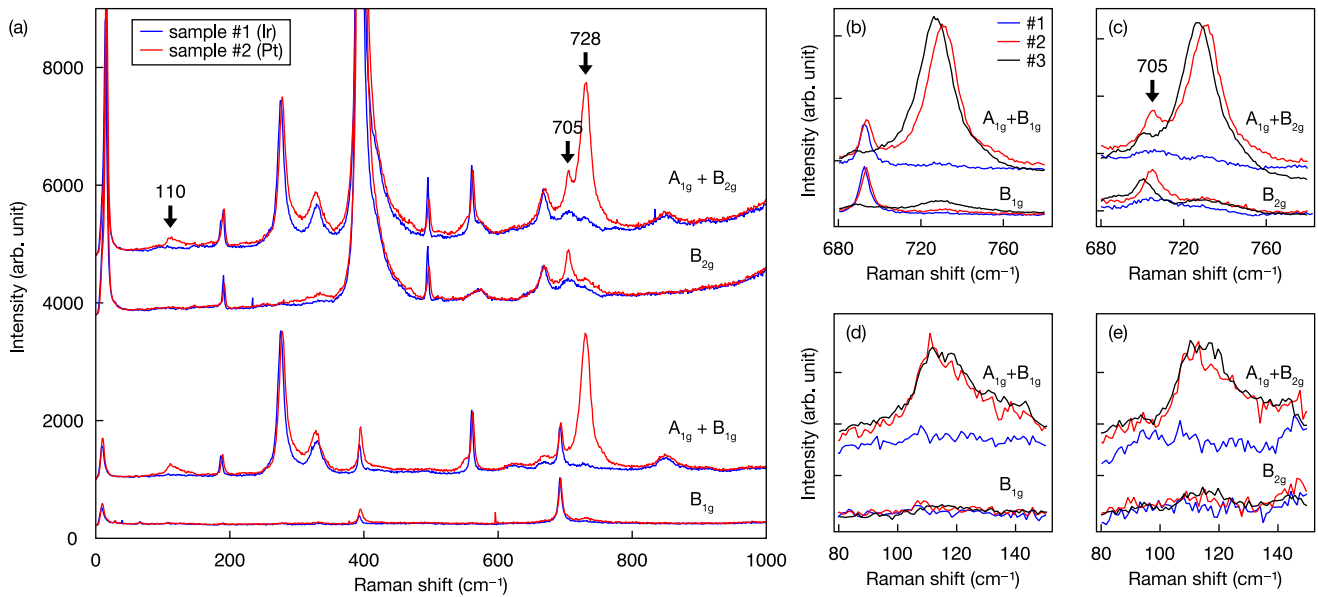}
\caption{\label{fig:1} (a) Raman spectra of Sr$_2$IrO$_4$ crystals grown in Ir (sample \#1, blue) and Pt (sample \#2, red) crucibles, taken at $T$ = 70 K in four different polarization configurations. (b)-(e) Detailed views that highlight the 705 cm$^{-1}$ $B_{2g}$ and 728 cm$^{-1}$ $A_{1g}$ peaks [(b) and (c)], and 110 cm$^{-1}$ $A_{1g}$ peak [(d) and (e)], which are suppressed in sample \#1. Black curves are Raman spectra of Pt-doped Sr$_2$IrO$_4$ crystals grown in an Ir crucible (sample \#3), taken at $T$ = 90 K.}
\end{figure*}

 In this paper, we show that growing Sr$_2$IrO$_4$ single crystals in a platinum (Pt) crucible introduces Pt impurities in the sample, which lead to many observable differences from those grown from an iridium (Ir) crucible. We assess the crystal quality using Raman spectroscopy, resonant X-ray diffraction (RXD), resonant inelastic X-ray scattering (RIXS), magnetization, resistivity, and optical THG-rotational anisotropy (RA) measurements. Remarkably, in Pt impurity-free samples, we find strong suppression of several peaks previously observed in the Raman spectra\cite{Ceti12, Gret16, Gret17}, and a large activation energy in good agreement with the theoretically calculated indirect band gap. Furthermore, we find no indication of the previously reported structural distortions. 
 Our result indicates that a small number of platinum impurities, below detection level of energy dispersive X-ray analysis (EDX), lead to manifest difference in the excitation spectra and the structure of Sr$_2$IrO$_4$, and possibly mask emergent phenomena that arise with carrier doping.


 Sr$_2$IrO$_4$ single crystals were grown using a self-flux method. IrO$_2$(99.99\%), SrCO$_3$(99.994\%), and SrCl$_2\cdot$6H$_2$O(99.9965\%) powders were mixed and ground thoroughly in an agate mortar, and placed in an iridium crucible enclosed by an outer alumina crucible. We used the same heating sequence as previously reported\cite{Sung16}, but the powders were mixed with a different molar ratio of reagents because the iridium crucible provides an iridium-rich environment\cite{supple}. Thin and shiny square-shaped single crystals with typical dimensions of 300 $\times$ 300 $\times$ 100 $\mu$m were extracted from the crucible after rinsing out with distilled water. 
 
 Raman spectroscopy was performed with a home-built setup equipped with a 633-nm He-Ne laser and a liquid-nitrogen-cooled CCD (Princeton Instruments). The elastic signal was removed by grating-based notch filters (Optigrate, BragGrate$^{\mathrm{TM}}$ Notch Filters). Sr$_2$IrO$_4$ crystals were mounted in an open cycle optical cryostat (Oxford Instruments) and cleaved to reveal fresh (001) surface. During the measurements, the nominal sample temperature was kept at 70 K. The laser power and spot size were 1.8 mW and 2 $\mu$m, respectively, and the resulting laser heating was about 50 K as determined from the Stokes to Anti-Stokes intensity ratio. All spectra were Bose-corrected. 
 
 THG-RA measurements were carried out by adopting the setup outlined in Ref.\cite{Hart15}. The phase mask was replaced by a pair of wedge prisms rotating at the speed of 1 Hz. The incident 1200 nm pulses were generated by an optical parametric amplifier (Light Conversion, Orpheus-F) powered by a femtosecond laser of 100 kHz repetition rate (Light Conversion, Pharos). The laser fluence was 0.6 mJ/cm$^2$, and the reflected third harmonic 400 nm signals were mapped onto a 2D electron multiplying CCD (Princeton Instruments). The crystal was cleaved at liquid nitrogen temperature and then immediately pumped down to pressure below 10$^{-6}$ torr. All THG-RA measurements were performed at room temperature.
 
 The in-plane resistivity was measured with a commercial physical property measurement system (Quantum Design). The standard six-probe method was used and the electrical current was applied to the $a$-axis on ion beam patterned Sr$_2$IrO$_4$ samples. The in-plane magnetization was measured with a commercial superconducting quantum interference device magnetometer (MPMS3-VSM, Quantum Design). 
 
 RXD measurements were performed at the 1C beamline of Pohang Light Source with the incident photon energy 11.215 keV and the focused beam spot size of 25(H) $\times$ 4(V) $\mu$m. We achieved an extremely low background in our RXD setup by using Si (844) crystal analyzer in back scattering geometry. This reduced the background intensity as low as one count per second (cps). RIXS spectra were obtained at the sector 27 RIXS beamline of Advanced Photon Source. X-rays were monochromatized to a bandwidth of 15 meV and focused to have a beam spot size of 40(H) $\times$ 10(V) $\mu$m. A Si (844) diced spherical analyzer with 2 inch radius was used and the achieved energy resolution at the iridium $L_3$ edge was 30 meV.



 Figure~\ref{fig:1} shows the comparison of the Raman spectra measured on single crystals grown in Ir (sample \#1) and Pt crucibles (sample \#2) at the liquid nitrogen temperature. Whereas most of the phonon modes of the samples \#1 and \#2 are in good agreement with each other and those in reported in earlier studies\cite{Ceti12, Gret16, Gret17}, the peaks at 110 cm$^{-1}$ and 728 cm$^{-1}$ in the $A_{1g}$ channel and at 705 cm$^{-1}$ in the $B_{2g}$ channel are strongly suppressed in sample \#1, as shown in Figs.~\ref{fig:1}(b)-(e). This result was reproduced in more than 10 crystals from other batches\cite{supple}. 
 To confirm the platinum impurity origin of these Raman peaks, we grew Sr$_2$IrO$_4$ crystals in an Ir crucible with extra PtO$_2$ powders(99.95\%) deliberately added (sample \#3). Black curves in Figs.~\ref{fig:1}(b)-(e) show that all three peaks (110 cm$^{-1}$, 705 cm$^{-1}$, 728 cm$^{-1}$) appear in sample \#3 as observed in sample \#2. Slight differences in the exact location of peaks and their intensity ratio are due to the different Pt impurity concentration and measurement temperature (70 K for \#2 and 90 K for \#3). Indeed, the 728 cm$^{-1}$ $A_{1g}$ peak shows the same peak location for sample \#2 and \#3 when measured at the same temperature\cite{supple}. Thus, we conclude that these three peaks originate from the platinum impurities and are not intrinsic to pure Sr$_2$IrO$_4$.

 \begin{figure}[t]
\includegraphics{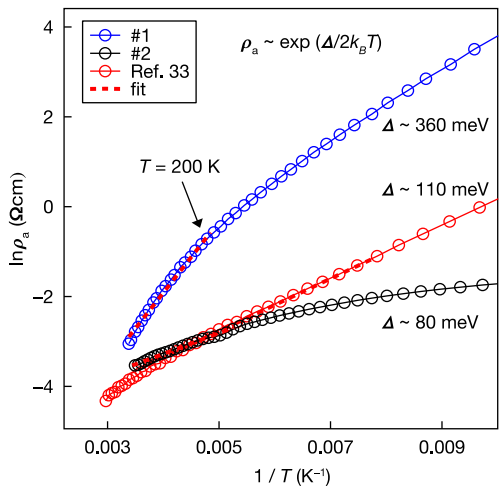}
\caption{\label{fig:2} An Arrhenius plot of the resistivity $\rho_a$ measured along the $a$-axis for sample \#1(blue circles), \#2(black circles), and that adopted from Ref.\cite{Ge11}(red circles). The energy gap $\Delta$ is extracted from the fitting (red dashed line) to the resistivity curve in the high temperature region.}
\end{figure}

From a representation analysis, it can be easily checked that three extra peaks become Raman active in the $A_{1g}$ channel as the lattice symmetry is lowered from $I4_1/acd$ to $I4_1/a$. As we shall see below, sample \#1 is confirmed to have $I4_1/acd$ symmetry. If the two $A_{1g}$ peaks (110 cm$^{-1}$ and 728 cm$^{-1}$) are phonon modes, their absence in sample \#1 can be interpreted as due to the lattice structural differences between sample \#1 and \#2. We note, however, the intensity of the peak at 728 cm$^{-1}$ in sample \#2 seem rather too large for a structural change induced by a small amount of Pt impurities below the detection level of EDX\cite{supple}.
 

 Alternatively, the peak may arise from in-gap electronic states created by the Pt impurities. As Pt has one more electron than Ir, Pt impurities can affect the transport properties. Figure~\ref{fig:2} shows the in-plane resistivity measured along the crystalline $a$-axis for sample \#1(blue circles) and \#2(black circles). The resistivity $\rho_a$ at high temperatures follows the activation law $\rho_a \sim \exp(\Delta/2k_BT)$, where $\Delta$ is the energy gap and $k_B$ is Boltzmann constant. The extracted $\Delta$ was about 360 meV, close to the theoretically calculated indirect band gap of about 300 meV based on the first-principles calculations\cite{Kim08, Mart17}. This $\Delta$ is significantly higher than previously reported values ranging from 80 meV to 110 meV\cite{Ge11, Wang15, Cao98, Korn10, Zocc14, Brou15, Piov16}, obtained in bulk crystals grown in a Pt crucible plotted as black and red circles in Fig.~\ref{fig:2}\cite{Ge11}. These observations suggest that some of the extra peaks in the Raman spectra found in crystals from Pt crucibles may be due to in-gap states from platinum impurities.
 
  
\begin{figure}[t]
\includegraphics{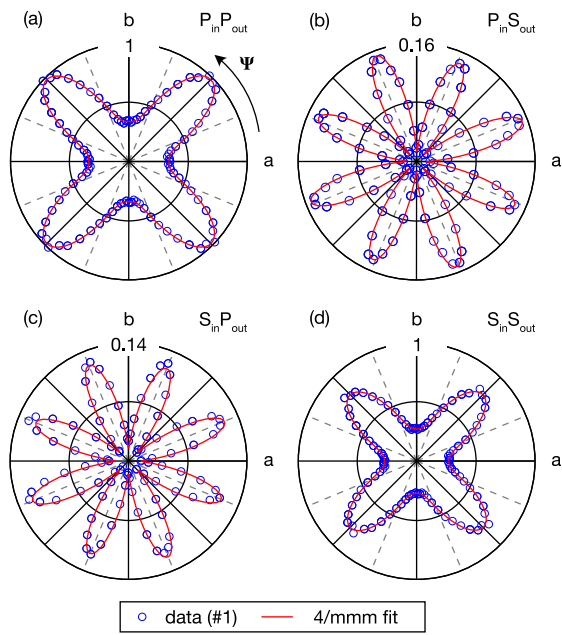}
\caption{\label{fig:3} THG-RA patterns of sample \#1 taken under (a) $PP$, (b) $PS$, (c) $SP$, (d) $SS$ at room temperature. All THG signals are normalized by the $PP$ trace. Red curves are the best fits using bulk electric dipole induced THG tensor of 4/$mmm$ point group. }
\end{figure}

 Next, we look into the issue of structural distortions recently reported for Sr$_2$IrO$_4$. The space group symmetry of Sr$_2$IrO$_4$ had long been believed to be $I4_1/acd$\cite{Craw94,Love12}, but recent works have instead suggested $I4_1/a$, based on the broken mirror symmetries observed in the THG-RA measurement\cite{Torc15}, and forbidden (1 0 $L$) peaks observed in the neutron diffraction experiment\cite{Dhit13, Ye13, Ye15}. We performed THG-RA measurements on our sample \#1 in four different polarization channels as shown in Fig.~\ref{fig:3}. The crystallographic axes were determined by the light polarization dependence of the phonon modes in our Raman spectra. We confirm that all four THG-RA patterns are well aligned with the crystallographic axes, and are perfectly fitted to the third harmonic generation susceptibility tensor in the electric dipole channel for $4/mmm$, the point group of $I4_1/acd$ space group. 
 
In the $4/mmm$ point group, the intensities for $PS$ and $SP$ geometries are proportional to $\left\vert \sin(4\Psi)\right\vert^2$, yielding a pattern with eight-fold rotation symmetry, as shown in Figs.~\ref{fig:3}(b) and (c), respectively. In the previous work, the lobes were tilted away from the principal axes and had different intensities, thus manifestly breaking the mirror symmetries $a\rightarrow -a$ and $b\rightarrow -b$, implying the absence of $c$ and $d$ glides from $I4_1/acd$\cite{Torc15,supple}. 


\begin{figure}[t]
\includegraphics{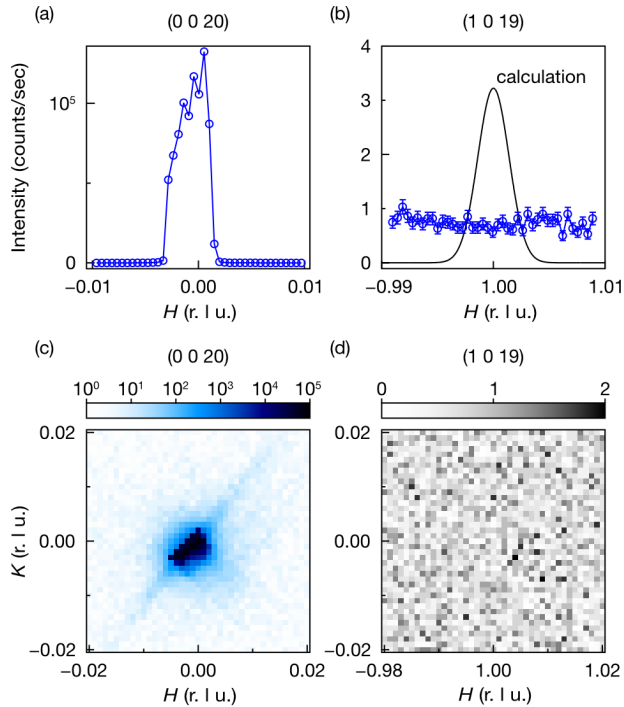}
\caption{\label{fig:4} RXD results for sample \#1, iridium crucible crystals. (a) Representative $H$ cut at (0 0 20). (b) Representative $H$ cut at (1 0 19). Black curve represents estimated theta rocking curve from structure factor calculation (see text). Error bar represents statistical error. (c), (d) Two dimensional $HK$ mapping for (0 0 20) and (1 0 19) reflections. }
\end{figure}

We have also searched for the previously reported forbidden reflections in our sample using RXD. We calculate the intensity expected for (1 0 19) in absolute units from the ratio of (1 0 19) to (0 0 20) based on their structure factors from the earlier structure refinement in $I4_1/a$\cite{Ye15}, and the intensity of (0 0 20) measured in our setup, as shown in Fig.~\ref{fig:4}(a). As only few cps is expected for (1 0 19), we used a Si (844) energy analyzer to suppress the background below one cps as shown in {Fig.~\ref{fig:4}(b). Our error bar of the order of $\sim$0.1 cps, an order of magnitude smaller than the expected intensity, allows us to confirm the absence of (1 0 19) reflection. To rule out the possibility of missing the peak due to a trivial sample misalignment, we performed a two-dimensional scan in ($H K$) plane as shown in Figs.~\ref{fig:4}(c) and (d). Our results from THG-RA and RXD comprehensively show that lattice structure of Sr$_2$IrO$_4$ is consistent with the $I4_1/acd$ space group.
 

\begin{figure}[b!]
\includegraphics{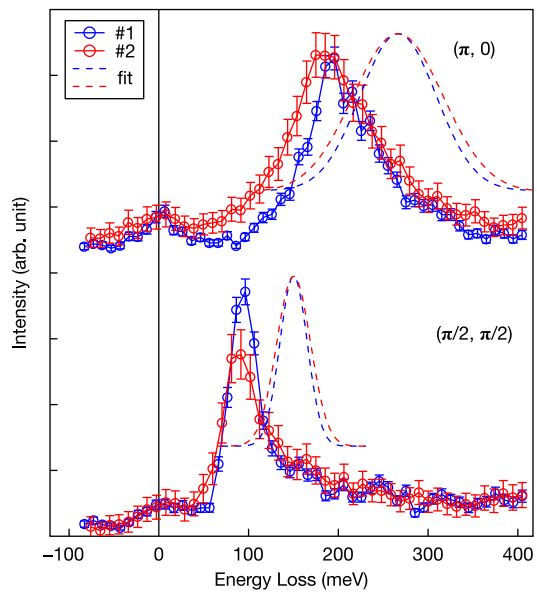}
\caption{\label{fig:5} High-resolution low energy RIXS spectra at two high symmetry points of ($\pi$, 0) and ($\pi$/2, $\pi$/2) grown in Ir (sample \#1, blue) and Pt (sample \#2, red) crucibles. The scattering angle was kept nearly 90$^\circ$ to suppress the elastic signal. The low energy features at 200 and 100 meV of ($\pi$, 0) and ($\pi$/2, $\pi$/2) correspond to single-magnon fitted with Gaussian curves (dashed line), respectively.}
\end{figure}

 The high quality of sample \#1 is further supported by RIXS. Figure~\ref{fig:5} compares the magnon spectra of sample \#1 and sample \#2 at the two high symmetry points at ($\pi$,0) and ($\pi/2$, $\pi/2$). It is clear that the magnon peaks in sample \#1 have significantly narrower peak widths. Indeed, the full widths at half maximum of the magnon peaks are 35.8 meV for sample \#1 and 44.2 meV for sample \#2 at ($\pi/2$, $\pi/2$), and 84.4 meV for sample \#1 and 99.5 meV for sample \#2 at ($\pi$, 0). Although our magnetometry measurements did not show any significant difference in the static magnetic response between the sample \#1 and \#2 \cite{supple}, the impurity effects are clearly visible in the dynamical spin structure factor. 

 In summary, we have shown that single crystal growth of Sr$_2$IrO$_4$ in Pt crucibles introduces Pt impurities, whose effects can be quite dramatic in the structural and physical properties of iridates. The impurities result in extra peaks in Raman spectra, in-gap states indicative of localized electrons bound to Pt impurities, and structural distortions breaking the glide symmetries. Given that magnon peaks become readily broadened with small amount of Pt impurities, novel quantum phases of matter such as the Kitaev spin liquid and topological Weyl semi-metal sought for in other iridates may require re-investigation using samples without Pt impurities. More importantly, absence of well-defined quasiparticles in metallic phases of electron and hole-doped Sr$_2$IrO$_4$ may be due to the impurities, as $in\:situ$ doping studies show prominent quasiparticle peaks indicative of coherent charge transport. Our studies suggest that reinvestigations of metallic phases of doped Sr$_2$IrO$_4$ in Pt impurity-free samples are necessary.

\begin{acknowledgements}

This project is supported by IBS-R014-A2. The use of the Advanced Photon Source at the Argonne National Laboratory was supported by the U. S. DOE under Contract No. DE-AC02-06CH11357. J. S. K. and H. W. S. acknowledge the support from the National Research Foundation of Korea (NRF)  (Grant No. NRF-2018R1A5A6075964).

\end{acknowledgements}



\bibliography{reference}

\begin{thebibliography}{40}%
\makeatletter
\providecommand \@ifxundefined [1]{%
 \@ifx{#1\undefined}
}%
\providecommand \@ifnum [1]{%
 \ifnum #1\expandafter \@firstoftwo
 \else \expandafter \@secondoftwo
 \fi
}%
\providecommand \@ifx [1]{%
 \ifx #1\expandafter \@firstoftwo
 \else \expandafter \@secondoftwo
 \fi
}%
\providecommand \natexlab [1]{#1}%
\providecommand \enquote  [1]{``#1''}%
\providecommand \bibnamefont  [1]{#1}%
\providecommand \bibfnamefont [1]{#1}%
\providecommand \citenamefont [1]{#1}%
\providecommand \href@noop [0]{\@secondoftwo}%
\providecommand \href [0]{\begingroup \@sanitize@url \@href}%
\providecommand \@href[1]{\@@startlink{#1}\@@href}%
\providecommand \@@href[1]{\endgroup#1\@@endlink}%
\providecommand \@sanitize@url [0]{\catcode `\\12\catcode `\$12\catcode
  `\&12\catcode `\#12\catcode `\^12\catcode `\_12\catcode `\%12\relax}%
\providecommand \@@startlink[1]{}%
\providecommand \@@endlink[0]{}%
\providecommand \url  [0]{\begingroup\@sanitize@url \@url }%
\providecommand \@url [1]{\endgroup\@href {#1}{\urlprefix }}%
\providecommand \urlprefix  [0]{URL }%
\providecommand \Eprint [0]{\href }%
\providecommand \doibase [0]{https://doi.org/}%
\providecommand \selectlanguage [0]{\@gobble}%
\providecommand \bibinfo  [0]{\@secondoftwo}%
\providecommand \bibfield  [0]{\@secondoftwo}%
\providecommand \translation [1]{[#1]}%
\providecommand \BibitemOpen [0]{}%
\providecommand \bibitemStop [0]{}%
\providecommand \bibitemNoStop [0]{.\EOS\space}%
\providecommand \EOS [0]{\spacefactor3000\relax}%
\providecommand \BibitemShut  [1]{\csname bibitem#1\endcsname}%
\let\auto@bib@innerbib\@empty
\bibitem [{\citenamefont {Kim}\ \emph {et~al.}(2008)\citenamefont {Kim},
  \citenamefont {Jin}, \citenamefont {Moon}, \citenamefont {Kim}, \citenamefont
  {Park}, \citenamefont {Leem}, \citenamefont {Yu}, \citenamefont {Noh},
  \citenamefont {Kim}, \citenamefont {Oh}, \citenamefont {Park}, \citenamefont
  {Durairaj}, \citenamefont {Cao},\ and\ \citenamefont {Rotenberg}}]{Kim08}%
  \BibitemOpen
  \bibfield  {author} {\bibinfo {author} {\bibfnamefont {B.~J.}\ \bibnamefont
  {Kim}}, \bibinfo {author} {\bibfnamefont {H.}~\bibnamefont {Jin}}, \bibinfo
  {author} {\bibfnamefont {S.~J.}\ \bibnamefont {Moon}}, \bibinfo {author}
  {\bibfnamefont {J.~Y.}\ \bibnamefont {Kim}}, \bibinfo {author} {\bibfnamefont
  {B.~G.}\ \bibnamefont {Park}}, \bibinfo {author} {\bibfnamefont {C.~S.}\
  \bibnamefont {Leem}}, \bibinfo {author} {\bibfnamefont {J.}~\bibnamefont
  {Yu}}, \bibinfo {author} {\bibfnamefont {T.~W.}\ \bibnamefont {Noh}},
  \bibinfo {author} {\bibfnamefont {C.}~\bibnamefont {Kim}}, \bibinfo {author}
  {\bibfnamefont {S.~J.}\ \bibnamefont {Oh}}, \bibinfo {author} {\bibfnamefont
  {J.~H.}\ \bibnamefont {Park}}, \bibinfo {author} {\bibfnamefont
  {V.}~\bibnamefont {Durairaj}}, \bibinfo {author} {\bibfnamefont
  {G.}~\bibnamefont {Cao}},\ and\ \bibinfo {author} {\bibfnamefont
  {E.}~\bibnamefont {Rotenberg}},\ }\bibfield  {title} {\bibinfo {title}
  {{{Novel $J_\textrm{eff}$=1/2 Mott State induced by Relativistic Spin-Orbit
  Coupling in Sr$_2$IrO$_4$}}},\ }\href
  {https://doi.org/10.1103/PhysRevLett.101.076402} {\bibfield  {journal}
  {\bibinfo  {journal} {Physical Review Letters}\ }\textbf {\bibinfo {volume}
  {101}},\ \bibinfo {pages} {76402} (\bibinfo {year} {2008})}\BibitemShut
  {NoStop}%
\bibitem [{\citenamefont {Kim}\ \emph {et~al.}(2009)\citenamefont {Kim},
  \citenamefont {Ohsumi}, \citenamefont {Komesu}, \citenamefont {Sakai},
  \citenamefont {Morita}, \citenamefont {Takagi},\ and\ \citenamefont
  {Arima}}]{Kim09}%
  \BibitemOpen
  \bibfield  {author} {\bibinfo {author} {\bibfnamefont {B.~J.}\ \bibnamefont
  {Kim}}, \bibinfo {author} {\bibfnamefont {H.}~\bibnamefont {Ohsumi}},
  \bibinfo {author} {\bibfnamefont {T.}~\bibnamefont {Komesu}}, \bibinfo
  {author} {\bibfnamefont {S.}~\bibnamefont {Sakai}}, \bibinfo {author}
  {\bibfnamefont {T.}~\bibnamefont {Morita}}, \bibinfo {author} {\bibfnamefont
  {H.}~\bibnamefont {Takagi}},\ and\ \bibinfo {author} {\bibfnamefont
  {T.}~\bibnamefont {Arima}},\ }\bibfield  {title} {\bibinfo {title}
  {{{Phase-Sensitive Observation of a Spin-Orbital Mott State in
  Sr$_2$IrO$_4$}}},\ }\href {https://doi.org/10.1126/science.1167106}
  {\bibfield  {journal} {\bibinfo  {journal} {Science}\ }\textbf {\bibinfo
  {volume} {323}},\ \bibinfo {pages} {1329} (\bibinfo {year}
  {2009})}\BibitemShut {NoStop}%
\bibitem [{\citenamefont {Jackeli}\ and\ \citenamefont
  {Khaliullin}(2009)}]{Jack09}%
  \BibitemOpen
  \bibfield  {author} {\bibinfo {author} {\bibfnamefont {G.}~\bibnamefont
  {Jackeli}}\ and\ \bibinfo {author} {\bibfnamefont {G.}~\bibnamefont
  {Khaliullin}},\ }\bibfield  {title} {\bibinfo {title} {Mott {{Insulators}} in
  the {{Strong Spin}}-{{Orbit Coupling Limit}}: {{From Heisenberg}} to a
  {{Quantum Compass}} and {{Kitaev Models}}},\ }\href
  {https://doi.org/10.1103/PhysRevLett.102.017205} {\bibfield  {journal}
  {\bibinfo  {journal} {Phys. Rev. Lett.}\ }\textbf {\bibinfo {volume} {102}},\
  \bibinfo {pages} {017205} (\bibinfo {year} {2009})}\BibitemShut {NoStop}%
\bibitem [{\citenamefont {Pesin}\ and\ \citenamefont {Balents}(2010)}]{Bale10}%
  \BibitemOpen
  \bibfield  {author} {\bibinfo {author} {\bibfnamefont {D.}~\bibnamefont
  {Pesin}}\ and\ \bibinfo {author} {\bibfnamefont {L.}~\bibnamefont
  {Balents}},\ }\bibfield  {title} {\bibinfo {title} {Mott physics and band
  topology in materials with strong spin–orbit interaction},\ }\href
  {https://doi.org/10.1038/nphys1606} {\bibfield  {journal} {\bibinfo
  {journal} {Nature Physics}\ }\textbf {\bibinfo {volume} {6}},\ \bibinfo
  {pages} {376} (\bibinfo {year} {2010})}\BibitemShut {NoStop}%
\bibitem [{\citenamefont {Witczak-Krempa}\ \emph {et~al.}(2013)\citenamefont
  {Witczak-Krempa}, \citenamefont {Chen}, \citenamefont {Kim},\ and\
  \citenamefont {Balents}}]{Witc13}%
  \BibitemOpen
  \bibfield  {author} {\bibinfo {author} {\bibfnamefont {W.}~\bibnamefont
  {Witczak-Krempa}}, \bibinfo {author} {\bibfnamefont {G.}~\bibnamefont
  {Chen}}, \bibinfo {author} {\bibfnamefont {Y.~B.}\ \bibnamefont {Kim}},\ and\
  \bibinfo {author} {\bibfnamefont {L.}~\bibnamefont {Balents}},\ }\bibfield
  {title} {\bibinfo {title} {{{Correlated Quantum Phenomena in the Strong
  Spin-Orbit Regime}}},\ }\href
  {https://doi.org/10.1146/annurev-conmatphys-020911-125138} {\bibfield
  {journal} {\bibinfo  {journal} {Annual Review of Condensed Matter Physics}\
  }\textbf {\bibinfo {volume} {5}},\ \bibinfo {pages} {57} (\bibinfo {year}
  {2013})}\BibitemShut {NoStop}%
\bibitem [{\citenamefont {Crawford}\ \emph {et~al.}(1994)\citenamefont
  {Crawford}, \citenamefont {Subramanian}, \citenamefont {Harlow},
  \citenamefont {Fernandez-Baca}, \citenamefont {Wang},\ and\ \citenamefont
  {Johnston}}]{Craw94}%
  \BibitemOpen
  \bibfield  {author} {\bibinfo {author} {\bibfnamefont {M.~K.}\ \bibnamefont
  {Crawford}}, \bibinfo {author} {\bibfnamefont {M.~A.}\ \bibnamefont
  {Subramanian}}, \bibinfo {author} {\bibfnamefont {R.~L.}\ \bibnamefont
  {Harlow}}, \bibinfo {author} {\bibfnamefont {J.~A.}\ \bibnamefont
  {Fernandez-Baca}}, \bibinfo {author} {\bibfnamefont {Z.~R.}\ \bibnamefont
  {Wang}},\ and\ \bibinfo {author} {\bibfnamefont {D.~C.}\ \bibnamefont
  {Johnston}},\ }\bibfield  {title} {\bibinfo {title} {{{Structural and
  Magnetic Studies of Sr$_2$IrO$_4$}}},\ }\href
  {https://doi.org/10.1103/PhysRevB.49.9198} {\bibfield  {journal} {\bibinfo
  {journal} {Phys. Rev. B}\ }\textbf {\bibinfo {volume} {49}},\ \bibinfo
  {pages} {9198} (\bibinfo {year} {1994})}\BibitemShut {NoStop}%
\bibitem [{\citenamefont {Cao}\ \emph {et~al.}(1998)\citenamefont {Cao},
  \citenamefont {Bolivar}, \citenamefont {McCall}, \citenamefont {Crow},\ and\
  \citenamefont {Guertin}}]{Cao98}%
  \BibitemOpen
  \bibfield  {author} {\bibinfo {author} {\bibfnamefont {G.}~\bibnamefont
  {Cao}}, \bibinfo {author} {\bibfnamefont {J.}~\bibnamefont {Bolivar}},
  \bibinfo {author} {\bibfnamefont {S.}~\bibnamefont {McCall}}, \bibinfo
  {author} {\bibfnamefont {J.~E.}\ \bibnamefont {Crow}},\ and\ \bibinfo
  {author} {\bibfnamefont {R.~P.}\ \bibnamefont {Guertin}},\ }\bibfield
  {title} {\bibinfo {title} {Weak ferromagnetism, metal-to-nonmetal transition,
  and negative differential resistivity in single-crystal {{Sr$_2$IrO$_4$}}},\
  }\href {https://doi.org/10.1103/PhysRevB.57.R11039} {\bibfield  {journal}
  {\bibinfo  {journal} {Phys. Rev. B}\ }\textbf {\bibinfo {volume} {57}},\
  \bibinfo {pages} {R11039} (\bibinfo {year} {1998})}\BibitemShut {NoStop}%
\bibitem [{\citenamefont {Wang}\ and\ \citenamefont {Senthil}(2011)}]{Wang11}%
  \BibitemOpen
  \bibfield  {author} {\bibinfo {author} {\bibfnamefont {F.~A.}\ \bibnamefont
  {Wang}}\ and\ \bibinfo {author} {\bibfnamefont {T.}~\bibnamefont {Senthil}},\
  }\bibfield  {title} {\bibinfo {title} {{{Twisted Hubbard Model for
  Sr$_2$IrO$_4$: Magnetism and Possible High Temperature Superconductivity}}},\
  }\href {https://doi.org/10.1103/PhysRevLett.106.136402} {\bibfield  {journal}
  {\bibinfo  {journal} {Physical Review Letters}\ }\textbf {\bibinfo {volume}
  {106}},\ \bibinfo {pages} {136402} (\bibinfo {year} {2011})}\BibitemShut
  {NoStop}%
\bibitem [{\citenamefont {Kim}\ \emph {et~al.}(2012)\citenamefont {Kim},
  \citenamefont {Casa}, \citenamefont {Upton}, \citenamefont {Gog},
  \citenamefont {Kim}, \citenamefont {Mitchell}, \citenamefont
  {Van~Veenendaal}, \citenamefont {Daghofer}, \citenamefont {Van Den~Brink},
  \citenamefont {Khaliullin},\ and\ \citenamefont {Kim}}]{Kim12}%
  \BibitemOpen
  \bibfield  {author} {\bibinfo {author} {\bibfnamefont {J.}~\bibnamefont
  {Kim}}, \bibinfo {author} {\bibfnamefont {D.}~\bibnamefont {Casa}}, \bibinfo
  {author} {\bibfnamefont {M.~H.}\ \bibnamefont {Upton}}, \bibinfo {author}
  {\bibfnamefont {T.}~\bibnamefont {Gog}}, \bibinfo {author} {\bibfnamefont
  {Y.~J.}\ \bibnamefont {Kim}}, \bibinfo {author} {\bibfnamefont {J.~F.}\
  \bibnamefont {Mitchell}}, \bibinfo {author} {\bibfnamefont {M.}~\bibnamefont
  {Van~Veenendaal}}, \bibinfo {author} {\bibfnamefont {M.}~\bibnamefont
  {Daghofer}}, \bibinfo {author} {\bibfnamefont {J.}~\bibnamefont {Van
  Den~Brink}}, \bibinfo {author} {\bibfnamefont {G.}~\bibnamefont
  {Khaliullin}},\ and\ \bibinfo {author} {\bibfnamefont {B.~J.}\ \bibnamefont
  {Kim}},\ }\bibfield  {title} {\bibinfo {title} {{{Magnetic Excitation Spectra
  of Sr$_2$IrO$_4$ Probed by Resonant Inelastic X-ray Scattering: Establishing
  Links to Cuprate Superconductors}}},\ }\href
  {https://doi.org/10.1103/PhysRevLett.108.177003} {\bibfield  {journal}
  {\bibinfo  {journal} {Physical Review Letters}\ }\textbf {\bibinfo {volume}
  {108}},\ \bibinfo {pages} {177003} (\bibinfo {year} {2012})}\BibitemShut
  {NoStop}%
\bibitem [{\citenamefont {Bertinshaw}\ \emph {et~al.}(2019)\citenamefont
  {Bertinshaw}, \citenamefont {Kim}, \citenamefont {Khaliullin},\ and\
  \citenamefont {Kim}}]{Bert19}%
  \BibitemOpen
  \bibfield  {author} {\bibinfo {author} {\bibfnamefont {J.}~\bibnamefont
  {Bertinshaw}}, \bibinfo {author} {\bibfnamefont {Y.~K.}\ \bibnamefont {Kim}},
  \bibinfo {author} {\bibfnamefont {G.}~\bibnamefont {Khaliullin}},\ and\
  \bibinfo {author} {\bibfnamefont {B.~J.}\ \bibnamefont {Kim}},\ }\bibfield
  {title} {\bibinfo {title} {{{Square Lattice Iridates}}},\ }\href
  {https://doi.org/10.1146/annurev-conmatphys-031218-013113} {\bibfield
  {journal} {\bibinfo  {journal} {Annual Review of Condensed Matter Physics}\
  }\textbf {\bibinfo {volume} {10}},\ \bibinfo {pages} {315} (\bibinfo {year}
  {2019})}\BibitemShut {NoStop}%
\bibitem [{\citenamefont {Watanabe}\ \emph {et~al.}(2013)\citenamefont
  {Watanabe}, \citenamefont {Shirakawa},\ and\ \citenamefont
  {Yunoki}}]{Wata13}%
  \BibitemOpen
  \bibfield  {author} {\bibinfo {author} {\bibfnamefont {H.}~\bibnamefont
  {Watanabe}}, \bibinfo {author} {\bibfnamefont {T.}~\bibnamefont
  {Shirakawa}},\ and\ \bibinfo {author} {\bibfnamefont {S.}~\bibnamefont
  {Yunoki}},\ }\bibfield  {title} {\bibinfo {title} {Monte {{Carlo Study}} of
  an {{Unconventional Superconducting Phase}} in {{Iridium Oxide
  $J_\textrm{eff}$}}=1/2 {{Mott Insulators Induced}} by {{Carrier Doping}}},\
  }\href {https://doi.org/10.1103/PhysRevLett.110.027002} {\bibfield  {journal}
  {\bibinfo  {journal} {Phys. Rev. Lett.}\ }\textbf {\bibinfo {volume} {110}},\
  \bibinfo {pages} {027002} (\bibinfo {year} {2013})}\BibitemShut {NoStop}%
\bibitem [{\citenamefont {Yang}\ \emph {et~al.}(2014)\citenamefont {Yang},
  \citenamefont {Wang}, \citenamefont {Liu}, \citenamefont {Chen},
  \citenamefont {Dai},\ and\ \citenamefont {Wang}}]{Yang14}%
  \BibitemOpen
  \bibfield  {author} {\bibinfo {author} {\bibfnamefont {Y.}~\bibnamefont
  {Yang}}, \bibinfo {author} {\bibfnamefont {W.-S.}\ \bibnamefont {Wang}},
  \bibinfo {author} {\bibfnamefont {J.-G.}\ \bibnamefont {Liu}}, \bibinfo
  {author} {\bibfnamefont {H.}~\bibnamefont {Chen}}, \bibinfo {author}
  {\bibfnamefont {J.-H.}\ \bibnamefont {Dai}},\ and\ \bibinfo {author}
  {\bibfnamefont {Q.-H.}\ \bibnamefont {Wang}},\ }\bibfield  {title} {\bibinfo
  {title} {{{Superconductivity in doped Sr$_2$IrO$_4$: A functional
  renormalization group study}}},\ }\href
  {https://doi.org/10.1103/PhysRevB.89.094518} {\bibfield  {journal} {\bibinfo
  {journal} {Physical Review B}\ }\textbf {\bibinfo {volume} {89}},\ \bibinfo
  {pages} {094518} (\bibinfo {year} {2014})}\BibitemShut {NoStop}%
\bibitem [{\citenamefont {Meng}\ \emph {et~al.}(2014)\citenamefont {Meng},
  \citenamefont {Kim},\ and\ \citenamefont {Kee}}]{Meng14}%
  \BibitemOpen
  \bibfield  {author} {\bibinfo {author} {\bibfnamefont {Z.~Y.}\ \bibnamefont
  {Meng}}, \bibinfo {author} {\bibfnamefont {Y.~B.}\ \bibnamefont {Kim}},\ and\
  \bibinfo {author} {\bibfnamefont {H.~Y.}\ \bibnamefont {Kee}},\ }\bibfield
  {title} {\bibinfo {title} {{{Odd-Parity triplet superconducting phase in
  multiorbital materials with a strong spin-orbit coupling: Application to
  doped Sr$_2$IrO$_4$}}},\ }\href
  {https://doi.org/10.1103/PhysRevLett.113.177003} {\bibfield  {journal}
  {\bibinfo  {journal} {Physical Review Letters}\ }\textbf {\bibinfo {volume}
  {113}},\ \bibinfo {pages} {177003} (\bibinfo {year} {2014})}\BibitemShut
  {NoStop}%
\bibitem [{\citenamefont {Kim}\ \emph {et~al.}(2014)\citenamefont {Kim},
  \citenamefont {Krupin}, \citenamefont {Denlinger}, \citenamefont {Bostwick},
  \citenamefont {Rotenberg}, \citenamefont {Zhao}, \citenamefont {Mitchell},
  \citenamefont {Allen},\ and\ \citenamefont {Kim}}]{Kim14}%
  \BibitemOpen
  \bibfield  {author} {\bibinfo {author} {\bibfnamefont {Y.~K.}\ \bibnamefont
  {Kim}}, \bibinfo {author} {\bibfnamefont {O.}~\bibnamefont {Krupin}},
  \bibinfo {author} {\bibfnamefont {J.~D.}\ \bibnamefont {Denlinger}}, \bibinfo
  {author} {\bibfnamefont {A.}~\bibnamefont {Bostwick}}, \bibinfo {author}
  {\bibfnamefont {E.}~\bibnamefont {Rotenberg}}, \bibinfo {author}
  {\bibfnamefont {Q.}~\bibnamefont {Zhao}}, \bibinfo {author} {\bibfnamefont
  {J.~F.}\ \bibnamefont {Mitchell}}, \bibinfo {author} {\bibfnamefont {J.~W.}\
  \bibnamefont {Allen}},\ and\ \bibinfo {author} {\bibfnamefont {B.~J.}\
  \bibnamefont {Kim}},\ }\bibfield  {title} {\bibinfo {title} {Fermi arcs in a
  doped pseudospin-1/2 heisenberg antiferromagnet},\ }\href
  {https://doi.org/10.1126/science.1251151} {\bibfield  {journal} {\bibinfo
  {journal} {Science}\ }\textbf {\bibinfo {volume} {345}},\ \bibinfo {pages}
  {187} (\bibinfo {year} {2014})}\BibitemShut {NoStop}%
\bibitem [{\citenamefont {Kim}\ \emph {et~al.}(2016)\citenamefont {Kim},
  \citenamefont {Sung}, \citenamefont {Denlinger},\ and\ \citenamefont
  {Kim}}]{Kim16}%
  \BibitemOpen
  \bibfield  {author} {\bibinfo {author} {\bibfnamefont {Y.~K.}\ \bibnamefont
  {Kim}}, \bibinfo {author} {\bibfnamefont {N.~H.}\ \bibnamefont {Sung}},
  \bibinfo {author} {\bibfnamefont {J.~D.}\ \bibnamefont {Denlinger}},\ and\
  \bibinfo {author} {\bibfnamefont {B.~J.}\ \bibnamefont {Kim}},\ }\bibfield
  {title} {\bibinfo {title} {Observation of a $d$-wave gap in electron-doped
  {{Sr$_2$IrO$_4$}}},\ }\href {https://doi.org/10.1038/nphys3503} {\bibfield
  {journal} {\bibinfo  {journal} {Nature Physics}\ }\textbf {\bibinfo {volume}
  {12}},\ \bibinfo {pages} {37} (\bibinfo {year} {2016})}\BibitemShut {NoStop}%
\bibitem [{\citenamefont {Yan}\ \emph {et~al.}(2015)\citenamefont {Yan},
  \citenamefont {Ren}, \citenamefont {Xu}, \citenamefont {Xie}, \citenamefont
  {Tao}, \citenamefont {Choi}, \citenamefont {Lee}, \citenamefont {Choi},
  \citenamefont {Zhang},\ and\ \citenamefont {Feng}}]{Yan15}%
  \BibitemOpen
  \bibfield  {author} {\bibinfo {author} {\bibfnamefont {Y.~J.}\ \bibnamefont
  {Yan}}, \bibinfo {author} {\bibfnamefont {M.~Q.}\ \bibnamefont {Ren}},
  \bibinfo {author} {\bibfnamefont {H.~C.}\ \bibnamefont {Xu}}, \bibinfo
  {author} {\bibfnamefont {B.~P.}\ \bibnamefont {Xie}}, \bibinfo {author}
  {\bibfnamefont {R.}~\bibnamefont {Tao}}, \bibinfo {author} {\bibfnamefont
  {H.~Y.}\ \bibnamefont {Choi}}, \bibinfo {author} {\bibfnamefont
  {N.}~\bibnamefont {Lee}}, \bibinfo {author} {\bibfnamefont {Y.~J.}\
  \bibnamefont {Choi}}, \bibinfo {author} {\bibfnamefont {T.}~\bibnamefont
  {Zhang}},\ and\ \bibinfo {author} {\bibfnamefont {D.~L.}\ \bibnamefont
  {Feng}},\ }\bibfield  {title} {\bibinfo {title} {{{Electron-Doped
  Sr$_2$IrO$_4$: An Analogue of Hole-Doped Cuprate Superconductors Demonstrated
  by Scanning Tunneling Microscopy}}},\ }\href
  {https://doi.org/10.1103/PhysRevX.5.041018} {\bibfield  {journal} {\bibinfo
  {journal} {Physical Review X}\ }\textbf {\bibinfo {volume} {5}},\ \bibinfo
  {pages} {041018} (\bibinfo {year} {2015})}\BibitemShut {NoStop}%
\bibitem [{\citenamefont {De~La~Torre}\ \emph {et~al.}(2015)\citenamefont
  {De~La~Torre}, \citenamefont {McKeown~Walker}, \citenamefont {Bruno},
  \citenamefont {Ricc$\acute{\textrm{o}}$}, \citenamefont {Wang}, \citenamefont
  {Gutierrez~Lezama}, \citenamefont {Scheerer}, \citenamefont {Giriat},
  \citenamefont {Jaccard}, \citenamefont {Berthod}, \citenamefont {Kim},
  \citenamefont {Hoesch}, \citenamefont {Hunter}, \citenamefont {Perry},
  \citenamefont {Tamai},\ and\ \citenamefont {Baumberger}}]{DeLa15}%
  \BibitemOpen
  \bibfield  {author} {\bibinfo {author} {\bibfnamefont {A.}~\bibnamefont
  {De~La~Torre}}, \bibinfo {author} {\bibfnamefont {S.}~\bibnamefont
  {McKeown~Walker}}, \bibinfo {author} {\bibfnamefont {F.~Y.}\ \bibnamefont
  {Bruno}}, \bibinfo {author} {\bibfnamefont {S.}~\bibnamefont
  {Ricc$\acute{\textrm{o}}$}}, \bibinfo {author} {\bibfnamefont
  {Z.}~\bibnamefont {Wang}}, \bibinfo {author} {\bibfnamefont {I.}~\bibnamefont
  {Gutierrez~Lezama}}, \bibinfo {author} {\bibfnamefont {G.}~\bibnamefont
  {Scheerer}}, \bibinfo {author} {\bibfnamefont {G.}~\bibnamefont {Giriat}},
  \bibinfo {author} {\bibfnamefont {D.}~\bibnamefont {Jaccard}}, \bibinfo
  {author} {\bibfnamefont {C.}~\bibnamefont {Berthod}}, \bibinfo {author}
  {\bibfnamefont {T.~K.}\ \bibnamefont {Kim}}, \bibinfo {author} {\bibfnamefont
  {M.}~\bibnamefont {Hoesch}}, \bibinfo {author} {\bibfnamefont {E.~C.}\
  \bibnamefont {Hunter}}, \bibinfo {author} {\bibfnamefont {R.~S.}\
  \bibnamefont {Perry}}, \bibinfo {author} {\bibfnamefont {A.}~\bibnamefont
  {Tamai}},\ and\ \bibinfo {author} {\bibfnamefont {F.}~\bibnamefont
  {Baumberger}},\ }\bibfield  {title} {\bibinfo {title} {{{Collapse of the Mott
  Gap and Emergence of a Nodal Liquid in Lightly Doped Sr$_2$IrO$_4$}}},\
  }\href {https://doi.org/10.1103/PhysRevLett.115.176402} {\bibfield  {journal}
  {\bibinfo  {journal} {Physical Review Letters}\ }\textbf {\bibinfo {volume}
  {115}},\ \bibinfo {pages} {176402} (\bibinfo {year} {2015})}\BibitemShut
  {NoStop}%
\bibitem [{\citenamefont {Cao}\ \emph {et~al.}(2016)\citenamefont {Cao},
  \citenamefont {Wang}, \citenamefont {Waugh}, \citenamefont {Reber},
  \citenamefont {Li}, \citenamefont {Zhou}, \citenamefont {Parham},
  \citenamefont {Park}, \citenamefont {Plumb}, \citenamefont {Rotenberg},
  \citenamefont {Bostwick}, \citenamefont {Denlinger}, \citenamefont {Qi},
  \citenamefont {Hermele}, \citenamefont {Cao},\ and\ \citenamefont
  {Dessau}}]{Cao16}%
  \BibitemOpen
  \bibfield  {author} {\bibinfo {author} {\bibfnamefont {Y.}~\bibnamefont
  {Cao}}, \bibinfo {author} {\bibfnamefont {Q.}~\bibnamefont {Wang}}, \bibinfo
  {author} {\bibfnamefont {J.~A.}\ \bibnamefont {Waugh}}, \bibinfo {author}
  {\bibfnamefont {T.~J.}\ \bibnamefont {Reber}}, \bibinfo {author}
  {\bibfnamefont {H.}~\bibnamefont {Li}}, \bibinfo {author} {\bibfnamefont
  {X.}~\bibnamefont {Zhou}}, \bibinfo {author} {\bibfnamefont {S.}~\bibnamefont
  {Parham}}, \bibinfo {author} {\bibfnamefont {S.~R.}\ \bibnamefont {Park}},
  \bibinfo {author} {\bibfnamefont {N.~C.}\ \bibnamefont {Plumb}}, \bibinfo
  {author} {\bibfnamefont {E.}~\bibnamefont {Rotenberg}}, \bibinfo {author}
  {\bibfnamefont {A.}~\bibnamefont {Bostwick}}, \bibinfo {author}
  {\bibfnamefont {J.~D.}\ \bibnamefont {Denlinger}}, \bibinfo {author}
  {\bibfnamefont {T.}~\bibnamefont {Qi}}, \bibinfo {author} {\bibfnamefont
  {M.~A.}\ \bibnamefont {Hermele}}, \bibinfo {author} {\bibfnamefont
  {G.}~\bibnamefont {Cao}},\ and\ \bibinfo {author} {\bibfnamefont {D.~S.}\
  \bibnamefont {Dessau}},\ }\bibfield  {title} {\bibinfo {title} {{{Hallmarks
  of the Mott-metal crossover in the hole-doped pseudospin-1/2 Mott insulator
  Sr$_2$IrO$_4$}}},\ }\href {https://doi.org/10.1038/ncomms11367} {\bibfield
  {journal} {\bibinfo  {journal} {Nature Communications}\ }\textbf {\bibinfo
  {volume} {7}},\ \bibinfo {pages} {1} (\bibinfo {year} {2016})}\BibitemShut
  {NoStop}%
\bibitem [{\citenamefont {Sun}\ \emph {et~al.}(2021)\citenamefont {Sun},
  \citenamefont {Guevara}, \citenamefont {Sykora}, \citenamefont {Pärschke},
  \citenamefont {Manna}, \citenamefont {Maljuk}, \citenamefont {Wurmehl},
  \citenamefont {van~den Brink}, \citenamefont {Büchner},\ and\ \citenamefont
  {Hess}}]{Sun21}%
  \BibitemOpen
  \bibfield  {author} {\bibinfo {author} {\bibfnamefont {Z.}~\bibnamefont
  {Sun}}, \bibinfo {author} {\bibfnamefont {J.~M.}\ \bibnamefont {Guevara}},
  \bibinfo {author} {\bibfnamefont {S.}~\bibnamefont {Sykora}}, \bibinfo
  {author} {\bibfnamefont {E.~M.}\ \bibnamefont {Pärschke}}, \bibinfo {author}
  {\bibfnamefont {K.}~\bibnamefont {Manna}}, \bibinfo {author} {\bibfnamefont
  {A.}~\bibnamefont {Maljuk}}, \bibinfo {author} {\bibfnamefont
  {S.}~\bibnamefont {Wurmehl}}, \bibinfo {author} {\bibfnamefont
  {J.}~\bibnamefont {van~den Brink}}, \bibinfo {author} {\bibfnamefont
  {B.}~\bibnamefont {Büchner}},\ and\ \bibinfo {author} {\bibfnamefont
  {C.}~\bibnamefont {Hess}},\ }\bibfield  {title} {\bibinfo {title} {{{Evidence
  for a percolative Mott insulator-metal transition in doped Sr$_2$IrO$_4$}}},\
  }\href {https://doi.org/10.1103/PhysRevResearch.3.023075} {\bibfield
  {journal} {\bibinfo  {journal} {Physical Review Research}\ }\textbf {\bibinfo
  {volume} {3}},\ \bibinfo {pages} {023075} (\bibinfo {year}
  {2021})}\BibitemShut {NoStop}%
\bibitem [{\citenamefont {Zhao}\ \emph {et~al.}(2021)\citenamefont {Zhao},
  \citenamefont {Porter}, \citenamefont {Chen}, \citenamefont {Wilson},
  \citenamefont {Wang},\ and\ \citenamefont {Zeljkovic}}]{Zhao21}%
  \BibitemOpen
  \bibfield  {author} {\bibinfo {author} {\bibfnamefont {H.}~\bibnamefont
  {Zhao}}, \bibinfo {author} {\bibfnamefont {Z.}~\bibnamefont {Porter}},
  \bibinfo {author} {\bibfnamefont {X.}~\bibnamefont {Chen}}, \bibinfo {author}
  {\bibfnamefont {S.~D.}\ \bibnamefont {Wilson}}, \bibinfo {author}
  {\bibfnamefont {Z.}~\bibnamefont {Wang}},\ and\ \bibinfo {author}
  {\bibnamefont {Zeljkovic}},\ }\bibfield  {title} {\bibinfo {title} {Imaging
  antiferromagnetic domain fluctuations and the effect of atomic scale disorder
  in a doped spin-orbit {{Mott}} insulator},\ }\href
  {https://doi.org/10.1126/sciadv.abi6468} {\bibfield  {journal} {\bibinfo
  {journal} {Science Advances}\ }\textbf {\bibinfo {volume} {7}},\ \bibinfo
  {pages} {eabi6468} (\bibinfo {year} {2021})}\BibitemShut {NoStop}%
\bibitem [{\citenamefont {Dhital}\ \emph {et~al.}(2013)\citenamefont {Dhital},
  \citenamefont {Hogan}, \citenamefont {Yamani}, \citenamefont {de~la Cruz},
  \citenamefont {Chen}, \citenamefont {Khadka}, \citenamefont {Ren},\ and\
  \citenamefont {Wilson}}]{Dhit13}%
  \BibitemOpen
  \bibfield  {author} {\bibinfo {author} {\bibfnamefont {C.}~\bibnamefont
  {Dhital}}, \bibinfo {author} {\bibfnamefont {T.}~\bibnamefont {Hogan}},
  \bibinfo {author} {\bibfnamefont {Z.}~\bibnamefont {Yamani}}, \bibinfo
  {author} {\bibfnamefont {C.}~\bibnamefont {de~la Cruz}}, \bibinfo {author}
  {\bibfnamefont {X.}~\bibnamefont {Chen}}, \bibinfo {author} {\bibfnamefont
  {S.}~\bibnamefont {Khadka}}, \bibinfo {author} {\bibfnamefont
  {Z.}~\bibnamefont {Ren}},\ and\ \bibinfo {author} {\bibfnamefont {S.~D.}\
  \bibnamefont {Wilson}},\ }\bibfield  {title} {\bibinfo {title} {Neutron
  scattering study of correlated phase behavior in {{Sr$_2$IrO$_4$}}},\ }\href
  {https://doi.org/10.1103/PhysRevB.87.144405} {\bibfield  {journal} {\bibinfo
  {journal} {Physical Review B}\ }\textbf {\bibinfo {volume} {87}},\ \bibinfo
  {pages} {144405} (\bibinfo {year} {2013})}\BibitemShut {NoStop}%
\bibitem [{\citenamefont {Ye}\ \emph {et~al.}(2013)\citenamefont {Ye},
  \citenamefont {Chi}, \citenamefont {Chakoumakos}, \citenamefont
  {Fernandez-Baca}, \citenamefont {Qi},\ and\ \citenamefont {Cao}}]{Ye13}%
  \BibitemOpen
  \bibfield  {author} {\bibinfo {author} {\bibfnamefont {F.}~\bibnamefont
  {Ye}}, \bibinfo {author} {\bibfnamefont {S.}~\bibnamefont {Chi}}, \bibinfo
  {author} {\bibfnamefont {B.~C.}\ \bibnamefont {Chakoumakos}}, \bibinfo
  {author} {\bibfnamefont {J.~A.}\ \bibnamefont {Fernandez-Baca}}, \bibinfo
  {author} {\bibfnamefont {T.}~\bibnamefont {Qi}},\ and\ \bibinfo {author}
  {\bibfnamefont {G.}~\bibnamefont {Cao}},\ }\bibfield  {title} {\bibinfo
  {title} {{{Magnetic and crystal structures of Sr$_2$IrO$_4$: A neutron
  diffraction study}}},\ }\href {https://doi.org/10.1103/PhysRevB.87.140406}
  {\bibfield  {journal} {\bibinfo  {journal} {Physical Review B}\ }\textbf
  {\bibinfo {volume} {87}},\ \bibinfo {pages} {140406(R)} (\bibinfo {year}
  {2013})}\BibitemShut {NoStop}%
\bibitem [{\citenamefont {Ye}\ \emph {et~al.}(2015)\citenamefont {Ye},
  \citenamefont {Wang}, \citenamefont {Hoffmann}, \citenamefont {Wang},
  \citenamefont {Chi}, \citenamefont {Matsuda}, \citenamefont {Chakoumakos},
  \citenamefont {Fernandez-Baca},\ and\ \citenamefont {Cao}}]{Ye15}%
  \BibitemOpen
  \bibfield  {author} {\bibinfo {author} {\bibfnamefont {F.}~\bibnamefont
  {Ye}}, \bibinfo {author} {\bibfnamefont {X.}~\bibnamefont {Wang}}, \bibinfo
  {author} {\bibfnamefont {C.}~\bibnamefont {Hoffmann}}, \bibinfo {author}
  {\bibfnamefont {J.}~\bibnamefont {Wang}}, \bibinfo {author} {\bibfnamefont
  {S.}~\bibnamefont {Chi}}, \bibinfo {author} {\bibfnamefont {M.}~\bibnamefont
  {Matsuda}}, \bibinfo {author} {\bibfnamefont {B.~C.}\ \bibnamefont
  {Chakoumakos}}, \bibinfo {author} {\bibfnamefont {J.~A.}\ \bibnamefont
  {Fernandez-Baca}},\ and\ \bibinfo {author} {\bibfnamefont {G.}~\bibnamefont
  {Cao}},\ }\bibfield  {title} {\bibinfo {title} {Structure symmetry
  determination and magnetic evolution in {{Sr$_2$Ir$_{1-x}$Rh$_x$O$_4$}}},\
  }\href {https://doi.org/10.1103/PhysRevB.92.201112} {\bibfield  {journal}
  {\bibinfo  {journal} {Physical Review B}\ }\textbf {\bibinfo {volume} {92}},\
  \bibinfo {pages} {201112(R)} (\bibinfo {year} {2015})}\BibitemShut {NoStop}%
\bibitem [{\citenamefont {Torchinsky}\ \emph {et~al.}(2015)\citenamefont
  {Torchinsky}, \citenamefont {Chu}, \citenamefont {Zhao}, \citenamefont
  {Perkins}, \citenamefont {Sizyuk}, \citenamefont {Qi}, \citenamefont {Cao},\
  and\ \citenamefont {Hsieh}}]{Torc15}%
  \BibitemOpen
  \bibfield  {author} {\bibinfo {author} {\bibfnamefont {D.~H.}\ \bibnamefont
  {Torchinsky}}, \bibinfo {author} {\bibfnamefont {H.}~\bibnamefont {Chu}},
  \bibinfo {author} {\bibfnamefont {L.}~\bibnamefont {Zhao}}, \bibinfo {author}
  {\bibfnamefont {N.~B.}\ \bibnamefont {Perkins}}, \bibinfo {author}
  {\bibfnamefont {Y.}~\bibnamefont {Sizyuk}}, \bibinfo {author} {\bibfnamefont
  {T.}~\bibnamefont {Qi}}, \bibinfo {author} {\bibfnamefont {G.}~\bibnamefont
  {Cao}},\ and\ \bibinfo {author} {\bibfnamefont {D.}~\bibnamefont {Hsieh}},\
  }\bibfield  {title} {\bibinfo {title} {{{Structural Distortion-Induced
  Magnetoelastic Locking in Sr$_2$IrO$_4$ revealed through Nonlinear Optical
  Harmonic Generation}}},\ }\href
  {https://doi.org/10.1103/PhysRevLett.114.096404} {\bibfield  {journal}
  {\bibinfo  {journal} {Physical Review Letters}\ }\textbf {\bibinfo {volume}
  {114}},\ \bibinfo {pages} {096404} (\bibinfo {year} {2015})}\BibitemShut
  {NoStop}%
\bibitem [{\citenamefont {Seyler}\ \emph {et~al.}(2020)\citenamefont {Seyler},
  \citenamefont {de~la Torre}, \citenamefont {Porter}, \citenamefont {Zoghlin},
  \citenamefont {Polski}, \citenamefont {Nguyen}, \citenamefont {Nadj-Perge},
  \citenamefont {Wilson},\ and\ \citenamefont {Hsieh}}]{Seyl20}%
  \BibitemOpen
  \bibfield  {author} {\bibinfo {author} {\bibfnamefont {K.~L.}\ \bibnamefont
  {Seyler}}, \bibinfo {author} {\bibfnamefont {A.}~\bibnamefont {de~la Torre}},
  \bibinfo {author} {\bibfnamefont {Z.}~\bibnamefont {Porter}}, \bibinfo
  {author} {\bibfnamefont {E.}~\bibnamefont {Zoghlin}}, \bibinfo {author}
  {\bibfnamefont {R.}~\bibnamefont {Polski}}, \bibinfo {author} {\bibfnamefont
  {M.}~\bibnamefont {Nguyen}}, \bibinfo {author} {\bibfnamefont
  {S.}~\bibnamefont {Nadj-Perge}}, \bibinfo {author} {\bibfnamefont {S.~D.}\
  \bibnamefont {Wilson}},\ and\ \bibinfo {author} {\bibfnamefont
  {D.}~\bibnamefont {Hsieh}},\ }\bibfield  {title} {\bibinfo {title}
  {Spin-orbit-enhanced magnetic surface second-harmonic generation in
  {{Sr$_2$IrO$_4$}}},\ }\href {https://doi.org/ARTN 201113
  10.1103/PhysRevB.102.201113} {\bibfield  {journal} {\bibinfo  {journal}
  {Physical Review B}\ }\textbf {\bibinfo {volume} {102}},\ \bibinfo {pages}
  {201113(R)} (\bibinfo {year} {2020})}\BibitemShut {NoStop}%
\bibitem [{\citenamefont {Sung}\ \emph {et~al.}(2016)\citenamefont {Sung},
  \citenamefont {Gretarsson}, \citenamefont {Proepper}, \citenamefont {Porras},
  \citenamefont {Le~Tacon}, \citenamefont {Boris}, \citenamefont {Keimer},\
  and\ \citenamefont {Kim}}]{Sung16}%
  \BibitemOpen
  \bibfield  {author} {\bibinfo {author} {\bibfnamefont {N.~H.}\ \bibnamefont
  {Sung}}, \bibinfo {author} {\bibfnamefont {H.}~\bibnamefont {Gretarsson}},
  \bibinfo {author} {\bibfnamefont {D.}~\bibnamefont {Proepper}}, \bibinfo
  {author} {\bibfnamefont {J.}~\bibnamefont {Porras}}, \bibinfo {author}
  {\bibfnamefont {M.}~\bibnamefont {Le~Tacon}}, \bibinfo {author}
  {\bibfnamefont {A.~V.}\ \bibnamefont {Boris}}, \bibinfo {author}
  {\bibfnamefont {B.}~\bibnamefont {Keimer}},\ and\ \bibinfo {author}
  {\bibfnamefont {B.~J.}\ \bibnamefont {Kim}},\ }\bibfield  {title} {\bibinfo
  {title} {Crystal growth and intrinsic magnetic behaviour of
  {{Sr$_2$IrO$_4$}}},\ }\href {https://doi.org/10.1080/14786435.2015.1134835}
  {\bibfield  {journal} {\bibinfo  {journal} {Philosophical Magazine}\ }\textbf
  {\bibinfo {volume} {96}},\ \bibinfo {pages} {413} (\bibinfo {year}
  {2016})}\BibitemShut {NoStop}%
\bibitem [{\citenamefont {Manna}\ \emph {et~al.}(2020)\citenamefont {Manna},
  \citenamefont {Aslan-Cansever}, \citenamefont {Maljuk}, \citenamefont
  {Wurmehl}, \citenamefont {Seiro},\ and\ \citenamefont
  {B$\ddot{\textrm{u}}$chner}}]{Mann20}%
  \BibitemOpen
  \bibfield  {author} {\bibinfo {author} {\bibfnamefont {K.}~\bibnamefont
  {Manna}}, \bibinfo {author} {\bibfnamefont {G.}~\bibnamefont
  {Aslan-Cansever}}, \bibinfo {author} {\bibfnamefont {A.}~\bibnamefont
  {Maljuk}}, \bibinfo {author} {\bibfnamefont {S.}~\bibnamefont {Wurmehl}},
  \bibinfo {author} {\bibfnamefont {S.}~\bibnamefont {Seiro}},\ and\ \bibinfo
  {author} {\bibfnamefont {B.}~\bibnamefont {B$\ddot{\textrm{u}}$chner}},\
  }\bibfield  {title} {\bibinfo {title} {Flux growth of
  {{Sr$_{n+1}$Ir$_n$O$_{3n+1}$}} ($n$ = 1, 2, $\infty$) crystals},\ }\href
  {https://doi.org/10.1016/j.jcrysgro.2020.125657} {\bibfield  {journal}
  {\bibinfo  {journal} {Journal of Crystal Growth}\ }\textbf {\bibinfo {volume}
  {540}},\ \bibinfo {pages} {125657} (\bibinfo {year} {2020})}\BibitemShut
  {NoStop}%
\bibitem [{\citenamefont {Cetin}\ \emph {et~al.}(2012)\citenamefont {Cetin},
  \citenamefont {Lemmens}, \citenamefont {Gnezdilov}, \citenamefont
  {Wulferding}, \citenamefont {Menzel}, \citenamefont {Takayama}, \citenamefont
  {Ohashi},\ and\ \citenamefont {Takagi}}]{Ceti12}%
  \BibitemOpen
  \bibfield  {author} {\bibinfo {author} {\bibfnamefont {M.~F.}\ \bibnamefont
  {Cetin}}, \bibinfo {author} {\bibfnamefont {P.}~\bibnamefont {Lemmens}},
  \bibinfo {author} {\bibfnamefont {V.}~\bibnamefont {Gnezdilov}}, \bibinfo
  {author} {\bibfnamefont {D.}~\bibnamefont {Wulferding}}, \bibinfo {author}
  {\bibfnamefont {D.}~\bibnamefont {Menzel}}, \bibinfo {author} {\bibfnamefont
  {T.}~\bibnamefont {Takayama}}, \bibinfo {author} {\bibfnamefont
  {K.}~\bibnamefont {Ohashi}},\ and\ \bibinfo {author} {\bibfnamefont
  {H.}~\bibnamefont {Takagi}},\ }\bibfield  {title} {\bibinfo {title}
  {{{Crossover from coherent to incoherent scattering in spin-orbit dominated
  Sr$_2$IrO$_4$}}},\ }\href {https://doi.org/10.1103/PhysRevB.85.195148}
  {\bibfield  {journal} {\bibinfo  {journal} {Physical Review B}\ }\textbf
  {\bibinfo {volume} {85}},\ \bibinfo {pages} {195148} (\bibinfo {year}
  {2012})}\BibitemShut {NoStop}%
\bibitem [{\citenamefont {Gretarsson}\ \emph {et~al.}(2016)\citenamefont
  {Gretarsson}, \citenamefont {Sung}, \citenamefont
  {H$\ddot{\textrm{o}}$ppner}, \citenamefont {Kim}, \citenamefont {Keimer},\
  and\ \citenamefont {Le~Tacon}}]{Gret16}%
  \BibitemOpen
  \bibfield  {author} {\bibinfo {author} {\bibfnamefont {H.}~\bibnamefont
  {Gretarsson}}, \bibinfo {author} {\bibfnamefont {N.~H.}\ \bibnamefont
  {Sung}}, \bibinfo {author} {\bibfnamefont {M.}~\bibnamefont
  {H$\ddot{\textrm{o}}$ppner}}, \bibinfo {author} {\bibfnamefont {B.~J.}\
  \bibnamefont {Kim}}, \bibinfo {author} {\bibfnamefont {B.}~\bibnamefont
  {Keimer}},\ and\ \bibinfo {author} {\bibfnamefont {M.}~\bibnamefont
  {Le~Tacon}},\ }\bibfield  {title} {\bibinfo {title} {{{Two-Magnon Raman
  Scattering and Pseudospin-Lattice Interactions in Sr$_2$IrO$_4$ and
  Sr$_3$Ir$_2$O$_7$}}},\ }\href
  {https://doi.org/10.1103/PhysRevLett.116.136401} {\bibfield  {journal}
  {\bibinfo  {journal} {Physical Review Letters}\ }\textbf {\bibinfo {volume}
  {116}},\ \bibinfo {pages} {136401} (\bibinfo {year} {2016})}\BibitemShut
  {NoStop}%
\bibitem [{\citenamefont {Gretarsson}\ \emph {et~al.}(2017)\citenamefont
  {Gretarsson}, \citenamefont {Sauceda}, \citenamefont {Sung}, \citenamefont
  {H$\ddot{\textrm{o}}$ppner}, \citenamefont {Minola}, \citenamefont {Kim},
  \citenamefont {Keimer},\ and\ \citenamefont {Le~Tacon}}]{Gret17}%
  \BibitemOpen
  \bibfield  {author} {\bibinfo {author} {\bibfnamefont {H.}~\bibnamefont
  {Gretarsson}}, \bibinfo {author} {\bibfnamefont {J.}~\bibnamefont {Sauceda}},
  \bibinfo {author} {\bibfnamefont {N.~H.}\ \bibnamefont {Sung}}, \bibinfo
  {author} {\bibfnamefont {M.}~\bibnamefont {H$\ddot{\textrm{o}}$ppner}},
  \bibinfo {author} {\bibfnamefont {M.}~\bibnamefont {Minola}}, \bibinfo
  {author} {\bibfnamefont {B.~J.}\ \bibnamefont {Kim}}, \bibinfo {author}
  {\bibfnamefont {B.}~\bibnamefont {Keimer}},\ and\ \bibinfo {author}
  {\bibfnamefont {M.}~\bibnamefont {Le~Tacon}},\ }\bibfield  {title} {\bibinfo
  {title} {Raman scattering study of vibrational and magnetic excitations in
  {{Sr$_{2-x}$La$_x$IrO$_4$}}},\ }\href {https://doi.org/ARTN 115138
  10.1103/PhysRevB.96.115138} {\bibfield  {journal} {\bibinfo  {journal}
  {Physical Review B}\ }\textbf {\bibinfo {volume} {96}},\ \bibinfo {pages}
  {115138} (\bibinfo {year} {2017})}\BibitemShut {NoStop}%
\bibitem [{sup()}]{supple}%
  \BibitemOpen
  \href@noop {} {}\bibinfo {note} {See Supplemental Material at [DOI] for
  details.}\BibitemShut {Stop}%
\bibitem [{\citenamefont {Harter}\ \emph {et~al.}(2015)\citenamefont {Harter},
  \citenamefont {Niu}, \citenamefont {Woss},\ and\ \citenamefont
  {Hsieh}}]{Hart15}%
  \BibitemOpen
  \bibfield  {author} {\bibinfo {author} {\bibfnamefont {J.~W.}\ \bibnamefont
  {Harter}}, \bibinfo {author} {\bibfnamefont {L.}~\bibnamefont {Niu}},
  \bibinfo {author} {\bibfnamefont {A.~J.}\ \bibnamefont {Woss}},\ and\
  \bibinfo {author} {\bibfnamefont {D.}~\bibnamefont {Hsieh}},\ }\bibfield
  {title} {\bibinfo {title} {High-speed measurement of rotational anisotropy
  nonlinear optical harmonic generation using position-sensitive detection},\
  }\href {https://doi.org/10.1364/OL.40.004671} {\bibfield  {journal} {\bibinfo
   {journal} {Optics Letters}\ }\textbf {\bibinfo {volume} {40}},\ \bibinfo
  {pages} {4671} (\bibinfo {year} {2015})}\BibitemShut {NoStop}%
\bibitem [{\citenamefont {Ge}\ \emph {et~al.}(2011)\citenamefont {Ge},
  \citenamefont {Qi}, \citenamefont {Korneta}, \citenamefont {De~Long},
  \citenamefont {Schlottmann}, \citenamefont {Crummett},\ and\ \citenamefont
  {Cao}}]{Ge11}%
  \BibitemOpen
  \bibfield  {author} {\bibinfo {author} {\bibfnamefont {M.}~\bibnamefont
  {Ge}}, \bibinfo {author} {\bibfnamefont {T.~F.}\ \bibnamefont {Qi}}, \bibinfo
  {author} {\bibfnamefont {O.~B.}\ \bibnamefont {Korneta}}, \bibinfo {author}
  {\bibfnamefont {D.~E.}\ \bibnamefont {De~Long}}, \bibinfo {author}
  {\bibfnamefont {P.}~\bibnamefont {Schlottmann}}, \bibinfo {author}
  {\bibfnamefont {W.~P.}\ \bibnamefont {Crummett}},\ and\ \bibinfo {author}
  {\bibfnamefont {G.}~\bibnamefont {Cao}},\ }\bibfield  {title} {\bibinfo
  {title} {Lattice-driven magnetoresistivity and metal-insulator transition in
  single-layered iridates},\ }\href
  {https://doi.org/10.1103/PhysRevB.84.100402} {\bibfield  {journal} {\bibinfo
  {journal} {Physical Review B}\ }\textbf {\bibinfo {volume} {84}},\ \bibinfo
  {pages} {100402(R)} (\bibinfo {year} {2011})}\BibitemShut {NoStop}%
\bibitem [{\citenamefont {Martins}\ \emph {et~al.}(2017)\citenamefont
  {Martins}, \citenamefont {Aichhorn},\ and\ \citenamefont
  {Biermann}}]{Mart17}%
  \BibitemOpen
  \bibfield  {author} {\bibinfo {author} {\bibfnamefont {C.}~\bibnamefont
  {Martins}}, \bibinfo {author} {\bibfnamefont {M.}~\bibnamefont {Aichhorn}},\
  and\ \bibinfo {author} {\bibfnamefont {S.}~\bibnamefont {Biermann}},\
  }\bibfield  {title} {\bibinfo {title} {Coulomb correlations in 4d and 5d
  oxides from first principles---or how spin-orbit materials choose their
  effective orbital degeneracies},\ }\href
  {https://doi.org/10.1088/1361-648X/aa648f} {\bibfield  {journal} {\bibinfo
  {journal} {J Phys Condens Matter}\ }\textbf {\bibinfo {volume} {29}},\
  \bibinfo {pages} {263001} (\bibinfo {year} {2017})}\BibitemShut {NoStop}%
\bibitem [{\citenamefont {Wang}\ \emph {et~al.}(2015)\citenamefont {Wang},
  \citenamefont {Seinige}, \citenamefont {Cao}, \citenamefont {Zhou},
  \citenamefont {Goodenough},\ and\ \citenamefont {Tsoi}}]{Wang15}%
  \BibitemOpen
  \bibfield  {author} {\bibinfo {author} {\bibfnamefont {C.}~\bibnamefont
  {Wang}}, \bibinfo {author} {\bibfnamefont {H.}~\bibnamefont {Seinige}},
  \bibinfo {author} {\bibfnamefont {G.}~\bibnamefont {Cao}}, \bibinfo {author}
  {\bibfnamefont {J.~S.}\ \bibnamefont {Zhou}}, \bibinfo {author}
  {\bibfnamefont {J.~B.}\ \bibnamefont {Goodenough}},\ and\ \bibinfo {author}
  {\bibfnamefont {M.}~\bibnamefont {Tsoi}},\ }\bibfield  {title} {\bibinfo
  {title} {{{Electrically tunable transport in the antiferromagnetic Mott
  insulator Sr$_2$IrO$_4$}}},\ }\href
  {https://doi.org/10.1103/PhysRevB.92.115136} {\bibfield  {journal} {\bibinfo
  {journal} {Physical Review B}\ }\textbf {\bibinfo {volume} {92}},\ \bibinfo
  {pages} {115136} (\bibinfo {year} {2015})}\BibitemShut {NoStop}%
\bibitem [{\citenamefont {Korneta}\ \emph {et~al.}(2010)\citenamefont
  {Korneta}, \citenamefont {Qi}, \citenamefont {Chikara}, \citenamefont
  {Parkin}, \citenamefont {De~Long}, \citenamefont {Schlottmann},\ and\
  \citenamefont {Cao}}]{Korn10}%
  \BibitemOpen
  \bibfield  {author} {\bibinfo {author} {\bibfnamefont {O.~B.}\ \bibnamefont
  {Korneta}}, \bibinfo {author} {\bibfnamefont {T.}~\bibnamefont {Qi}},
  \bibinfo {author} {\bibfnamefont {S.}~\bibnamefont {Chikara}}, \bibinfo
  {author} {\bibfnamefont {S.}~\bibnamefont {Parkin}}, \bibinfo {author}
  {\bibfnamefont {L.~E.}\ \bibnamefont {De~Long}}, \bibinfo {author}
  {\bibfnamefont {P.}~\bibnamefont {Schlottmann}},\ and\ \bibinfo {author}
  {\bibfnamefont {G.}~\bibnamefont {Cao}},\ }\bibfield  {title} {\bibinfo
  {title} {{{Electron-doped {Sr}$_2${IrO}$_{4-\delta}$($0\leq\delta\leq0.04$):
  Evolution of a disordered ${J}_{\textrm{eff}}$=$\frac{1}{2}$ Mott insulator
  into an exotic metallic state}}},\ }\href
  {https://doi.org/10.1103/PhysRevB.82.115117} {\bibfield  {journal} {\bibinfo
  {journal} {Physical Review B}\ }\textbf {\bibinfo {volume} {82}},\ \bibinfo
  {pages} {115117} (\bibinfo {year} {2010})}\BibitemShut {NoStop}%
\bibitem [{\citenamefont {Zocco}\ \emph {et~al.}(2014)\citenamefont {Zocco},
  \citenamefont {Hamlin}, \citenamefont {White}, \citenamefont {Kim},
  \citenamefont {Jeffries}, \citenamefont {Weir}, \citenamefont {Vohra},
  \citenamefont {Allen},\ and\ \citenamefont {Maple}}]{Zocc14}%
  \BibitemOpen
  \bibfield  {author} {\bibinfo {author} {\bibfnamefont {D.~A.}\ \bibnamefont
  {Zocco}}, \bibinfo {author} {\bibfnamefont {J.~J.}\ \bibnamefont {Hamlin}},
  \bibinfo {author} {\bibfnamefont {B.~D.}\ \bibnamefont {White}}, \bibinfo
  {author} {\bibfnamefont {B.~J.}\ \bibnamefont {Kim}}, \bibinfo {author}
  {\bibfnamefont {J.~R.}\ \bibnamefont {Jeffries}}, \bibinfo {author}
  {\bibfnamefont {S.~T.}\ \bibnamefont {Weir}}, \bibinfo {author}
  {\bibfnamefont {Y.~K.}\ \bibnamefont {Vohra}}, \bibinfo {author}
  {\bibfnamefont {J.~W.}\ \bibnamefont {Allen}},\ and\ \bibinfo {author}
  {\bibfnamefont {M.~B.}\ \bibnamefont {Maple}},\ }\bibfield  {title} {\bibinfo
  {title} {Persistent non-metallic behavior in {Sr}$_2${IrO}$_4$ and
  {Sr}$_3${Ir}$_2${O}$_7$ at high pressures},\ }\href
  {https://doi.org/10.1088/0953-8984/26/25/255603} {\bibfield  {journal}
  {\bibinfo  {journal} {Journal of Physics: Condensed Matter}\ }\textbf
  {\bibinfo {volume} {26}},\ \bibinfo {pages} {255603} (\bibinfo {year}
  {2014})}\BibitemShut {NoStop}%
\bibitem [{\citenamefont {Brouet}\ \emph {et~al.}(2015)\citenamefont {Brouet},
  \citenamefont {Mansart}, \citenamefont {Perfetti}, \citenamefont {Piovera},
  \citenamefont {Vobornik}, \citenamefont {Le~F{\`e}vre}, \citenamefont
  {Bertran}, \citenamefont {Riggs}, \citenamefont {Shapiro}, \citenamefont
  {Giraldo-Gallo},\ and\ \citenamefont {Fisher}}]{Brou15}%
  \BibitemOpen
  \bibfield  {author} {\bibinfo {author} {\bibfnamefont {V.}~\bibnamefont
  {Brouet}}, \bibinfo {author} {\bibfnamefont {J.}~\bibnamefont {Mansart}},
  \bibinfo {author} {\bibfnamefont {L.}~\bibnamefont {Perfetti}}, \bibinfo
  {author} {\bibfnamefont {C.}~\bibnamefont {Piovera}}, \bibinfo {author}
  {\bibfnamefont {I.}~\bibnamefont {Vobornik}}, \bibinfo {author}
  {\bibfnamefont {P.}~\bibnamefont {Le~F{\`e}vre}}, \bibinfo {author}
  {\bibfnamefont {F.}~\bibnamefont {Bertran}}, \bibinfo {author} {\bibfnamefont
  {S.~C.}\ \bibnamefont {Riggs}}, \bibinfo {author} {\bibfnamefont {M.~C.}\
  \bibnamefont {Shapiro}}, \bibinfo {author} {\bibfnamefont {P.}~\bibnamefont
  {Giraldo-Gallo}},\ and\ \bibinfo {author} {\bibfnamefont {I.~R.}\
  \bibnamefont {Fisher}},\ }\bibfield  {title} {\bibinfo {title} {Transfer of
  spectral weight across the gap of {Sr}$_2${IrO}$_4$ induced by {La} doping},\
  }\href {https://doi.org/10.1103/PhysRevB.92.081117} {\bibfield  {journal}
  {\bibinfo  {journal} {Physical Review B}\ }\textbf {\bibinfo {volume} {92}},\
  \bibinfo {pages} {081117} (\bibinfo {year} {2015})}\BibitemShut {NoStop}%
\bibitem [{\citenamefont {Piovera}\ \emph {et~al.}(2016)\citenamefont
  {Piovera}, \citenamefont {Brouet}, \citenamefont {Papalazarou}, \citenamefont
  {Caputo}, \citenamefont {Marsi}, \citenamefont {Taleb-Ibrahimi},
  \citenamefont {Kim},\ and\ \citenamefont {Perfetti}}]{Piov16}%
  \BibitemOpen
  \bibfield  {author} {\bibinfo {author} {\bibfnamefont {C.}~\bibnamefont
  {Piovera}}, \bibinfo {author} {\bibfnamefont {V.}~\bibnamefont {Brouet}},
  \bibinfo {author} {\bibfnamefont {E.}~\bibnamefont {Papalazarou}}, \bibinfo
  {author} {\bibfnamefont {M.}~\bibnamefont {Caputo}}, \bibinfo {author}
  {\bibfnamefont {M.}~\bibnamefont {Marsi}}, \bibinfo {author} {\bibfnamefont
  {A.}~\bibnamefont {Taleb-Ibrahimi}}, \bibinfo {author} {\bibfnamefont
  {B.~J.}\ \bibnamefont {Kim}},\ and\ \bibinfo {author} {\bibfnamefont
  {L.}~\bibnamefont {Perfetti}},\ }\bibfield  {title} {\bibinfo {title}
  {Time-resolved photoemission of {Sr}$_2${IrO}$_4$},\ }\href
  {https://doi.org/10.1103/PhysRevB.93.241114} {\bibfield  {journal} {\bibinfo
  {journal} {Physical Review B}\ }\textbf {\bibinfo {volume} {93}},\ \bibinfo
  {pages} {241114} (\bibinfo {year} {2016})}\BibitemShut {NoStop}%
\bibitem [{\citenamefont {Lovesey}\ \emph {et~al.}(2012)\citenamefont
  {Lovesey}, \citenamefont {Khalyavin}, \citenamefont {Manuel}, \citenamefont
  {Chapon}, \citenamefont {Cao},\ and\ \citenamefont {Qi}}]{Love12}%
  \BibitemOpen
  \bibfield  {author} {\bibinfo {author} {\bibfnamefont {S.~W.}\ \bibnamefont
  {Lovesey}}, \bibinfo {author} {\bibfnamefont {D.~D.}\ \bibnamefont
  {Khalyavin}}, \bibinfo {author} {\bibfnamefont {P.}~\bibnamefont {Manuel}},
  \bibinfo {author} {\bibfnamefont {L.~C.}\ \bibnamefont {Chapon}}, \bibinfo
  {author} {\bibfnamefont {G.}~\bibnamefont {Cao}},\ and\ \bibinfo {author}
  {\bibfnamefont {T.~F.}\ \bibnamefont {Qi}},\ }\bibfield  {title} {\bibinfo
  {title} {Magnetic symmetries in neutron and resonant x-ray bragg diffraction
  patterns of four iridium oxides},\ }\href
  {https://doi.org/10.1088/0953-8984/24/49/496003} {\bibfield  {journal}
  {\bibinfo  {journal} {J Phys Condens Matter}\ }\textbf {\bibinfo {volume}
  {24}},\ \bibinfo {pages} {496003} (\bibinfo {year} {2012})}\BibitemShut
  {NoStop}%
\end{thebibliography}%

\end{document}